\documentclass[10pt,aps,hidelinks,twocolumn, prl, superscriptaddress,float fix]{revtex4-2}
\usepackage{graphicx}
\usepackage{dcolumn}
\usepackage{orcidlink}
\usepackage{caption}

\usepackage[T1]{fontenc}
\usepackage{subcaption}
\usepackage{csquotes}
\usepackage{float}
\usepackage{amsfonts}
\usepackage{amsmath}
\usepackage{amsmath,amssymb}
\usepackage[italicdiff]{physics}
\usepackage{bm}
\usepackage{mathtools}
\usepackage{enumitem}
\usepackage{tensor}
\usepackage{wrapfig}
\usepackage{footnote}
\usepackage{here, comment}
\usepackage{bm}
\begin{document}


\title{Timelike Entanglement Signatures of Ergodicity and Spectral Chaos}

\author{Rathindra Nath Das\,\orcidlink{0000-0002-4766-7705}}
\email{das.rathindranath@uni-wuerzburg.de}
\affiliation{Department of Particle Physics and Astrophysics, Weizmann Institute of Science, Rehovot 7610001, Israel}
\affiliation{MIT Center for Theoretical Physics—a Leinweber Institute, Massachusetts Institute of Technology,
Cambridge, MA 02139, USA}
\affiliation{Institute for Theoretical Physics and Astrophysics and Würzburg-Dresden Cluster of Excellence ctd.qmat, Julius-Maximilians-Universität Würzburg, Am Hubland, 97074 Würzburg, Germany}

\author{Arnab Kundu\,\orcidlink{0000-0002-1994-3346}}
\email{arnab.kundu@saha.ac.in}
\affiliation{Theory division, Saha Institute of Nuclear Physics, A CI of Homi Bhabha National Institute, 1/AF, Bidhannagar, Kolkata 700064, India}

\author{Nemai Chandra Sarkar\,\orcidlink{0009-0009-0275-7705}}
\email{nemaichandra.sarkar@saha.ac.in}
\affiliation{Theory division, Saha Institute of Nuclear Physics, A CI of Homi Bhabha National Institute, 1/AF, Bidhannagar, Kolkata 700064, India}

\date{\today}

\begin{abstract}

We investigate timelike entanglement measures derived from the spacetime density kernel in the Rosenzweig-Porter model and show that they sharply diagnose both eigenvector ergodicity and spectral chaos. For several Hilbert-space bipartitions, we compute the second Tsallis entropy, the entanglement imagitivity that quantifies non-Hermiticity, and Schatten-norm diagnostics of the kernel. The imagitivity and Frobenius norm exhibit rapid growth and high late-time plateaus in the ergodic regime, are suppressed in the localized regime, and show intermediate behavior in the fractal phase. The real part of the second Tsallis entropy displays a spectral form factor-like dip-ramp-plateau throughout the chaotic window and a suppressed ramp in the localized regime. We further introduce a kernel negativity, defined as the negative spectral weight of the Hermitian part of the kernel. This negativity equals the trace-norm distance to the set of positive semidefinite operators and the maximal witnessable negative quasiprobability, and its time-averaged value decreases across the ergodic-fractal-localized crossover in close correspondence with the fractal dimension.

\end{abstract}

\maketitle

\paragraph*{Introduction---} The study of quantum chaos occupies a central position at the interface of quantum information, high-energy physics, and many-body dynamics. The spread of quantum information from the fast scrambling of black holes \cite{Sekino_2008} to the breakdown of thermalization in many-body localized phases is typically diagnosed using distinct tools. Early-time scrambling is captured by out-of-time-ordered correlators \cite{Larkin1969-wj,Maldacena:2015waa}, while late-time quantum ergodicity and chaos are revealed through spectral statistics and the dip-ramp-plateau structure of the spectral form factor \cite{mehta2004random,Bohigas:1983er,Haake:2010fgh}. A unified language that treats temporal and spatial correlations on comparable footing, however, remains limited. 

Recent works have extended entanglement beyond spatial bipartitions into the time domain \cite{Milekhin:2025ycm,Guo:2025dtq, Nakata:2020luh,Mollabashi:2020yie, Mollabashi:2021xsd,Doi:2022iyj, Doi:2023zaf, Caputa:2024gve,Kawamoto:2025oko,Buscemi:2013xlk,Parzygnat:2022pax, Horsman:2017nqa, Fullwood:2022rjd,Afrasiar:2024lsi,Afrasiar:2024ldn,Afrasiar:2025eam,He:2024jog,Narayan:2023ebn,Nanda:2025tid,Caputa:2025ugm,Guo:2025ase}. Timelike entanglement entropy, defined via analytic continuation of standard entanglement measures \cite{Milekhin:2025ycm}, treats distinct time slices as entangled subsystems and naturally leads to complex pseudo-entropies associated with transition matrices \cite{Nakata:2020luh,Mollabashi:2020yie,Mollabashi:2021xsd,Caputa:2024gve}. From a broader perspective, the generalized spacetime density kernel \cite{Das:2025fcd} packages multi-time correlation functions into a single operator. It simultaneously encodes early-time scrambling and late-time spectral rigidity within the same object, connecting conventional chaos diagnostics to the entanglement measures.

Amid these developments, the Rosenzweig--Porter (RP) ensemble \cite{PhysRev.120.1698,Buijsman:2023ips,Bogomolny:2018nyh,De_Tomasi_2019,PhysRevResearch.2.043346,PhysRevB.103.104205,Khaymovich_2021,Buijsman:2021xbi,PhysRevB.106.094204,Venturelli_2023,PhysRevB.108.L060203,Roy_2025,Kutlin_2024} provides an ideal testbed because it interpolates between an ergodic phase, a localized phase, and an intermediate non-ergodic extended fractal regime \cite{Kravtsov_2015,De_Tomasi_2019,Altland_1997,von_Soosten_2018}, while simultaneously exhibiting a crossover from GOE-like to Poisson-like spectral statistics \cite{Pandey1995,BrezinHikami1996,PhysRevE.58.400}. Although inverse participation ratios and level statistics resolve these phases as static properties \cite{Bogomolny:2018nyh,Buijsman:2021xbi}, they do not directly characterize how temporal correlations and timelike entanglement develop under real-time evolution.

In this Letter, we introduce a timelike-entanglement approach to probe fractality and spectral chaos in the RP model. We evaluate the time evolution of a multiqubit system under the RP Hamiltonian and construct a spacetime density kernel for distinct Hilbert-space bipartitions. 
In the ergodic phase, timelike entanglement measures grow rapidly at early times and saturate at high long-time values, reflecting fast information spreading. In particular, the Frobenius norm of the spacetime kernel, which quantifies the overall strength of two-time correlations, exhibits a steep initial rise followed by a maximal saturation value in the ergodic regime. The fractal phase displays intermediate behavior: the kernel norm grows more slowly and saturates at a lower plateau, indicating partial yet constrained exploration of configuration space. These trends are mirrored by other Schatten-norm diagnostics of the spacetime kernel. The imagitivity, which quantifies non-Hermiticity, is largest in the ergodic regime and is progressively reduced across the fractal phase, becoming minimal in the localized regime.

In addition, we show that the same spacetime-kernel framework captures spectral chaos. The real part of the second Tsallis entropy exhibits a spectral form factor-like dip-ramp-plateau structure that closely tracks the Haar averaged two-point correlator $\overline{F}_2(t,0)$ introduced in Ref.~\cite{Das:2025fcd}. In the RP model, the ramp in both the second Tsallis entropy and $\overline{F}_2(t,0)$ persists throughout the chaotic window ($\gamma\lesssim 2$) and is strongly suppressed for $\gamma>2$, consistent with the loss of long-range spectral rigidity in the localized regime.

Another central result of our work is the introduction of a \emph{kernel negativity} as a quantitative marker of nonclassical temporal correlations. We define the kernel negativity as the total negative spectral weight of the Hermitian part of the spacetime density kernel. Equivalently, it is obtained by diagonalizing the symmetrized kernel and summing the magnitudes of its negative eigenvalues; this quantity equals the trace-norm distance of the symmetrized kernel from the set of positive semidefinite operators and corresponds to the maximal negative quasiprobability that can be witnessed by a positive operator-valued measure (POVM).
We find that this kernel negativity is highly sensitive to the RP phase: it is largest in the ergodic regime, takes intermediate values in the fractal phase, and is strongly suppressed in the localized regime where the kernel becomes effectively diagonal. We note that the variation of kernel negativity across the phase diagram closely mirrors the behavior of the eigenstate fractal dimension. By leveraging these timelike entanglement-based observables, our work bridges quantum information and quantum chaos diagnostics, unifying eigenvector ergodicity and spectral chaos within a single spacetime-kernel framework.

\begin{figure*}[hbtp]
    \centering
     \begin{subfigure}[b]{0.30\textwidth}
     \centering
         \includegraphics[width=\textwidth]{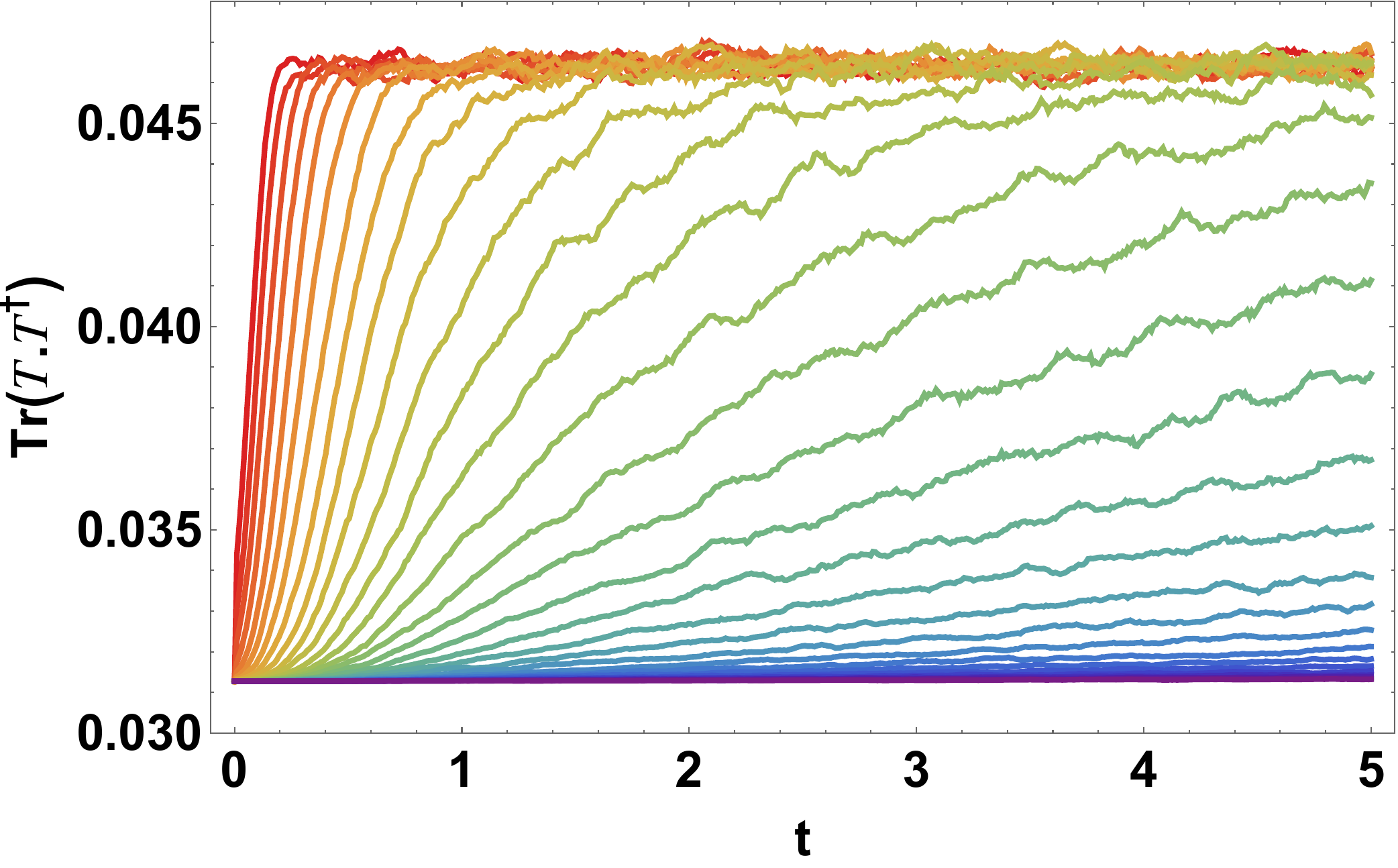}
         \caption{}
         \label{fig:ttbar}
     \end{subfigure}
     \hfill
     \begin{subfigure}[b]{0.32\textwidth}
         \centering
         \includegraphics[width=\textwidth]{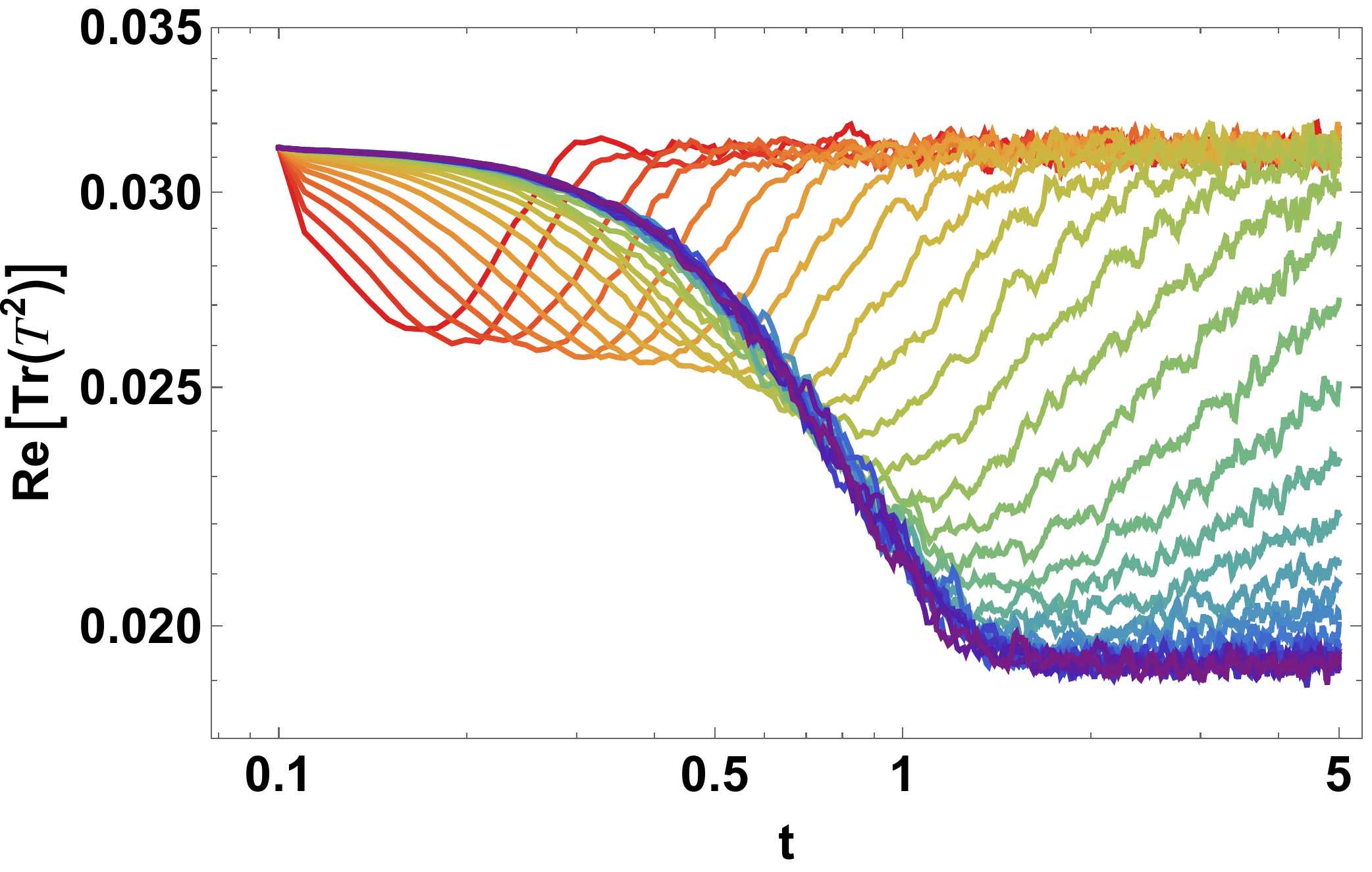}
         \caption{}
         \label{fig:tsq}
     \end{subfigure}
     \hfill
     \begin{subfigure}[b]{0.32\textwidth}
         \centering
         \includegraphics[width=\textwidth]{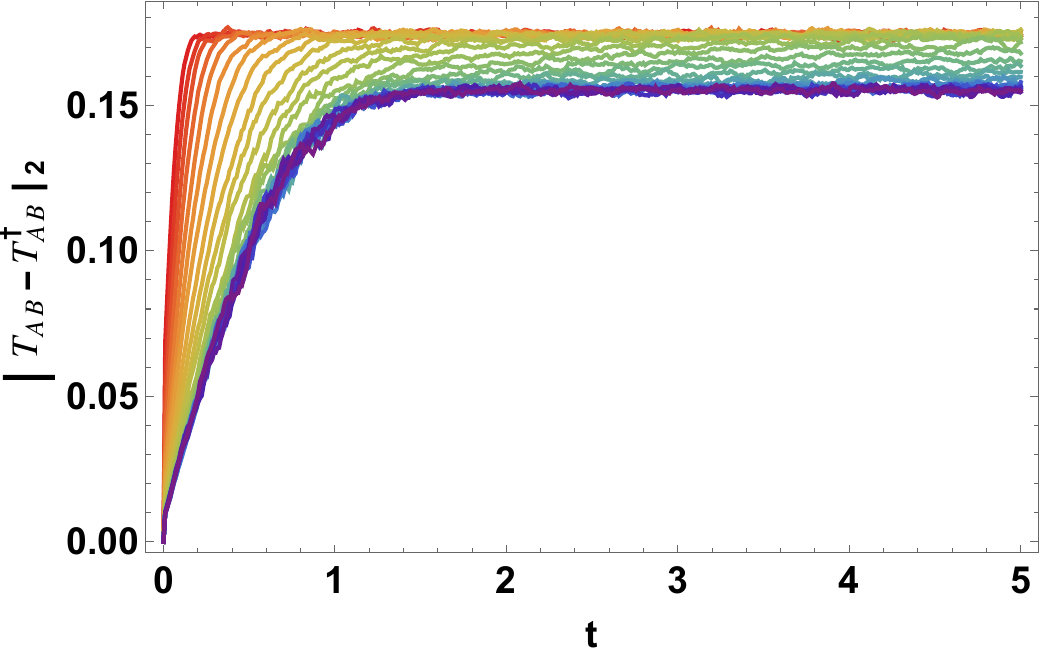}
         \caption{}
         \label{fig:schntab}
     \end{subfigure}
     \\
     \begin{subfigure}[b]{0.3\textwidth}
     \centering
         \includegraphics[width=\textwidth]{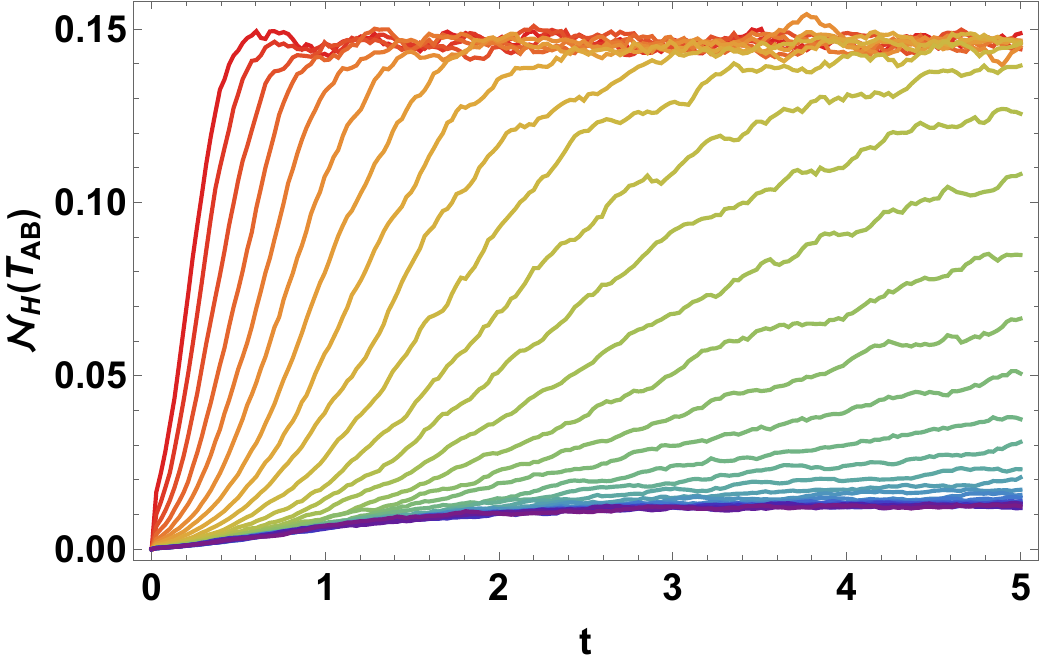}
         \caption{}
         \label{fig:negtab}
     \end{subfigure}
     \hfill
     \begin{subfigure}[b]{0.3\textwidth}
     \centering
         \includegraphics[width=\textwidth]{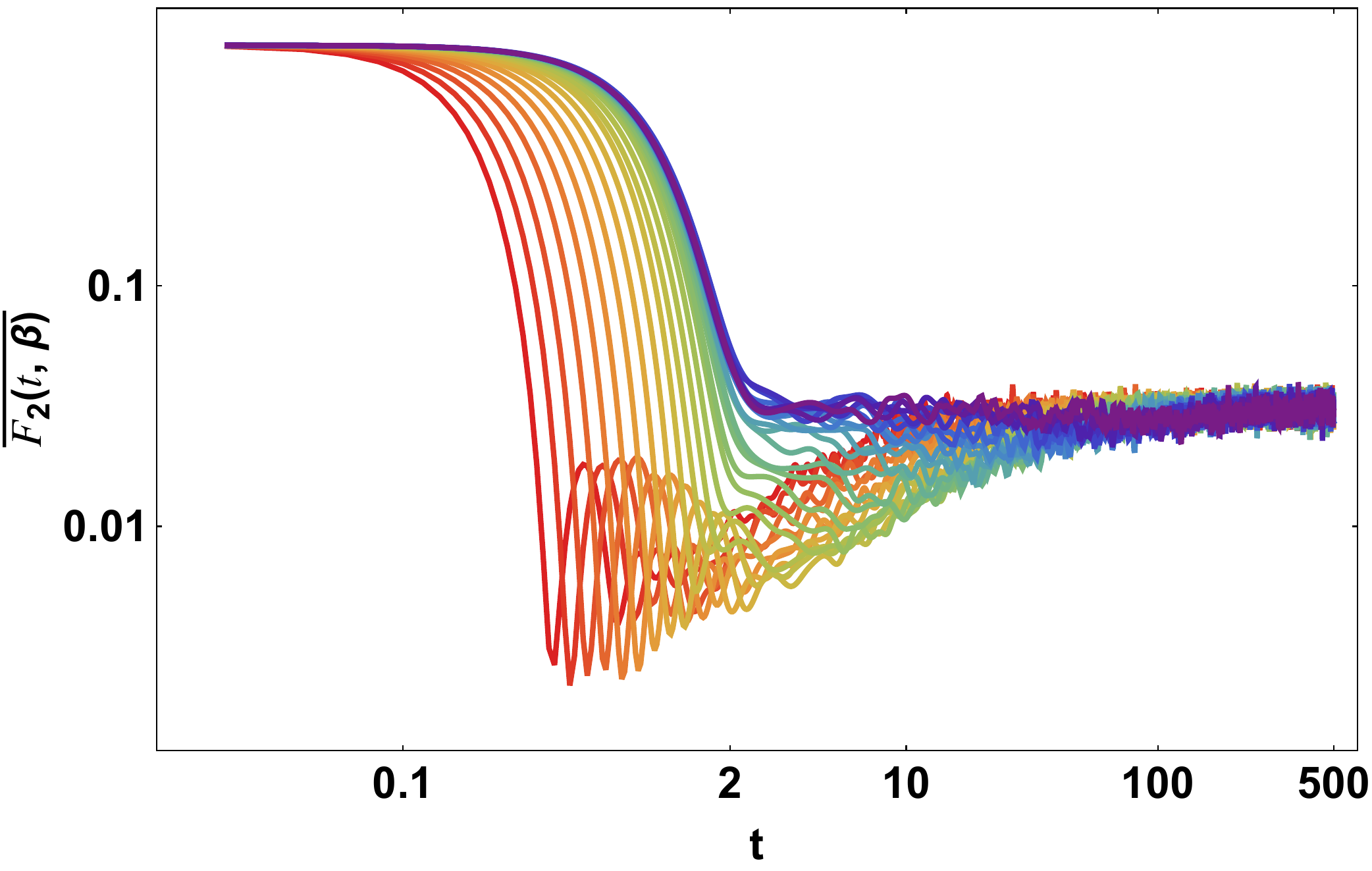}
         \caption{}
         \label{fig:f2bar}
     \end{subfigure}
     \hfill
     \begin{subfigure}[b]{0.35\textwidth}
         \centering
         \includegraphics[width=\textwidth]{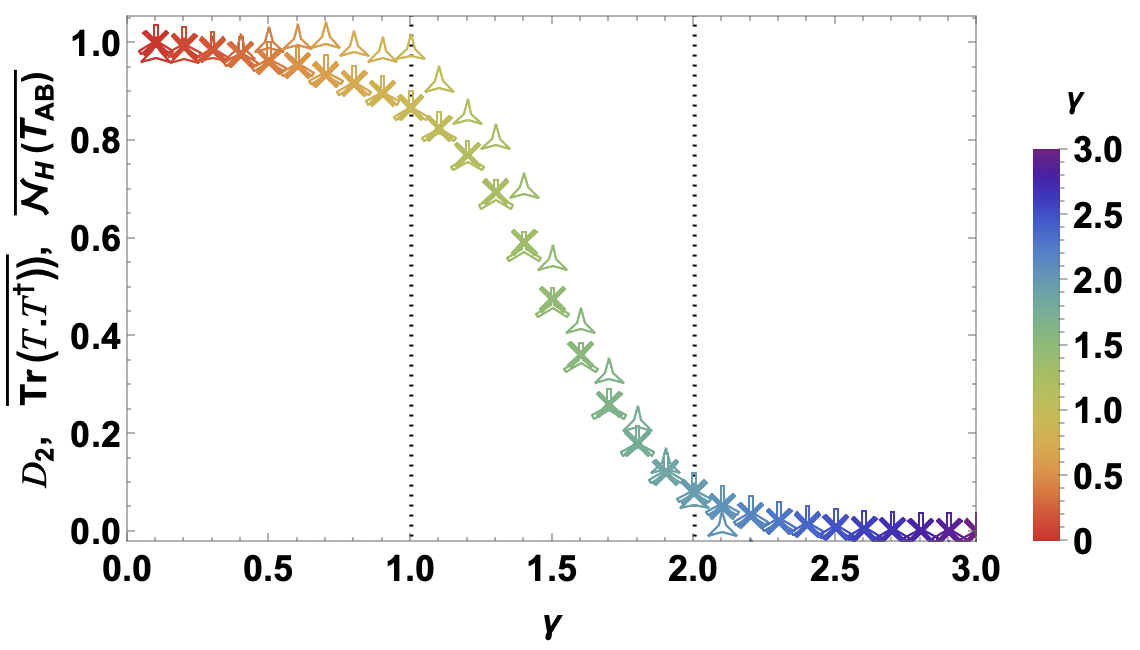}
         \caption{}
         \label{fig:D2_ttbar_ntavg}
     \end{subfigure}
     
        \caption{(a) Frobenius norm $\|T_{AB}(t)\|_2^2$ vs $t$: faster growth and higher plateau at small $\gamma$, suppressed at large $\gamma$, with the fractal regime interpolating.
(b) $\mathrm{Re}\,\mathrm{Tr}[T_{AB}(t)^2]$ vs $t$: SFF-like dip--ramp--plateau in the ergodic regime; ramp weakens in the fractal regime and is absent when localized.
(c) $2$-imagitivity $\mathcal I_2(t)$ vs $t$: non-Hermiticity grows and saturates most strongly for small $\gamma$ and is  reduced for large $\gamma$.
(d) Kernel negativity $\mathcal N_H(t)$ vs $t$: negative spectral weight develops at late times; its plateau decreases with increasing $\gamma$. (e) $\overline{F_2(t,\beta=0)}$ vs $t$: $\overline{F_2(t,0)}$ behaves as the normalised spectral form factor (SFF) showing dip--ramp--plateau in the ergodic regime; ramp weakens in the fractal regime and vanishes when localized.
(f) $D_2$ (star) and rescaled time-averaged $\overline{\mathcal N_H(T_{AB})}$ (diagonal-cross) and $\overline{\|T_{AB}\|_2^2}$ (triple-cross) vs $\gamma$ (dashed lines at $\gamma=1,2$): $\overline{\mathcal N_H(T_{AB})}$ and $\overline{\|T_{AB}\|_2^2}$ decrease in tandem with the loss of eigenvector ergodicity---tracking the same ergodic--fractal--localized crossover signaled by $D_2$.
}
\label{fig:1}
\end{figure*}
\paragraph*{RP model and timelike entanglement---} 
We consider a quantum system of $M$ qubits, so that the Hilbert space has dimension $N=2^M$. The system is bipartitioned into two subsystems $A$ and $B$. The dynamics is governed by the Rosenzweig-Porter ensemble \cite{Kravtsov_2015}, a random-matrix Hamiltonian that interpolates between ergodic and localized regimes while retaining a preferred basis. In our implementation, the Hamiltonian takes the form $H = H_0 + \frac{\varepsilon}{N^{\gamma/2}}V$, where $H_0$ is diagonal with independently distributed on-site energies and $V$ is drawn from the Gaussian orthogonal ensemble, modeling all-to-all couplings. The control parameter $\gamma$ tunes both eigenstate structure and spectral statistics: for $0 < \gamma < 1$, the eigenstates are fully ergodic in the basis of $H_0$ and the level statistics are of Wigner-Dyson (GOE) type; for $2 < \gamma$, the eigenstates become nearly localized and the level statistics cross over to Poisson. The intermediate regime $1<\gamma<2$ hosts a non-ergodic extended fractal phase, characterized by eigenstates occupying a subextensive fraction of the Hilbert space. Quantitatively, this regime is described by the fractal dimension $D_2$ \cite{Bogomolny_2018}, defined through the scaling of the inverse participation ratio $\text{IPR}_2 \sim N^{-D_2}$, with $D_2 = 1$ in the ergodic phase, $D_2 = 0$ in the localized phase, and $0 < D_2 < 1$ in the fractal regime.

To probe timelike correlations between subsystem $A$ at time $0$ and subsystem $B$ at time $t$, we package all two-time Wightman correlators into a single operator $T_{AB}(t)$ acting on $\mathcal{H}_A\otimes\mathcal{H}_B$ \footnote{Because the Rosenzweig--Porter ensemble is a zero-dimensional quantum model without spatial structure, our use of “spacetime density kernel” packages two-time correlators between subsystems $A$ at time $0$ and $B$ at time $t$ defined by qubit bipartitions.}
. Concretely, $T_{AB}(t)$ is defined by the trace-pairing identity that holds for arbitrary probes $O_A$ on $A$ and $O_B$ on $B$,
\begin{equation}
\mathrm{Tr}\!\left[T_{AB}(t)\,(O_A\otimes O_B)\right]
= \mathrm{Tr}\!\left(\rho\, O_A\, U_t^\dagger\, O_B\, U_t\right),
\label{eq:tab_def_pairing}
\end{equation}
where $U_t=e^{-iHt}$ and $\rho$ is the initial state. Unless stated otherwise, we take $\rho=\ket{\psi_0}\bra{\psi_0}$ with $\ket{\psi_0}$ a random pure state on the $M$-qubit Hilbert space.
This is the operator-basis formulation of the spacetime density matrix construction: in a matrix-unit basis $E^{A}_{ij}=|i\rangle\langle j|$ and $E^{B}_{kl}=|k\rangle\langle l|$, one forms the “generalized response tensor” $C_{ij;kl}(t)=\mathrm{Tr}\!\left(\rho\,\widehat E^{A}_{ij}\,U_t^\dagger\,\widehat E^{B}_{kl}\,U_t\right)$ and reconstructs
\begin{equation}
T_{AB}(t)=\sum_{i,j,k,l} C_{ij;kl}(t)\,E^A_{ji}\otimes E^B_{lk},
\label{eq:tab_basis_recon}
\end{equation}
where the flipped indices ensure $\mathrm{Tr}(E_{ji}X)=X_{ij}$.  In contrast to an ordinary reduced density matrix, $T_{AB}(t)$ need not be Hermitian. Nevertheless, $T_{AB}(t)$ has unit trace, $\mathrm{Tr}\,T_{AB}(t)=1$. 

Following ~\cite{Milekhin:2025ycm}, we monitor the Tsallis entropy
\begin{equation}
Z_n(t)\equiv \mathrm{Tr}\!\left[T_{AB}(t)^n\right],
\label{eq:Zn_def}
\end{equation}
which can be complex in general. Here we focus on $n=2$.  We further characterize $T_{AB}(t)$ through Schatten norms. The Schatten $p$-norm is defined as $\|M\|_p=(\mathrm{Tr}(M^\dagger M)^{p/2})^{1/p}$. A particularly useful quantity is the Frobenius norm,
\begin{equation}
\|T_{AB}(t)\|_2^2=\mathrm{Tr}\!\left[T_{AB}(t)\,T^\dagger_{AB}(t)\right],
\label{eq:frob_def}
\end{equation}
which we compute alongside $Z_2(t)$. To quantify non-Hermiticity, we use the entanglement \emph{$p$-imagitivity}
\begin{equation}
\mathcal{I}_p(t)\equiv \|T_{AB}(t)-T^\dagger_{AB}(t)\|_p,
\qquad (\text{we take }p=2),
\label{eq:imag_def}
\end{equation}
introduced in Ref.~\cite{Milekhin:2025ycm}. The case $p=2$ is singled out because it tightly controls causal influence: $\mathcal{I}_2(t)$ upper bounds time-dependent commutators and is nonzero iff signaling from $A$ to $B$ is possible. Algebraically, the 2-imagitivity relates the Frobenius norm and the second Tsallis entropy via
$\mathcal{I}_2(t)^2
=\|T_{AB}(t)-T^\dagger_{AB}(t)\|_2^2
=2\,\mathrm{Tr}\!\left[T_{AB}(t)T^\dagger_{AB}(t)\right]-2\,\mathrm{Re}\,\mathrm{Tr}\!\left[T_{AB}(t)^2\right]$,
so $\|T_{AB}\|_2^2$ and $\mathrm{Re}\,Z_2$ jointly diagnose the growth of non-Hermiticity.

To compare the above kernel based measures to a standard spectral diagnostic, we also track the
Haar-averaged two-point function introduced in Ref.~\cite{Das:2025fcd}.
Using the equal-spacing finite-temperature regularization, define $y$ by $y^{4}\propto e^{-\beta H}$ and consider the regulated response tensor $ C_{ij;kl}^{(\beta)}(t)=\operatorname{Tr}\!\big(y\,E_{ij}\,y\,U_t^\dagger\,y\,E_{kl}\,y\,U_t\big)$.
Averaging over probes with the normalized Haar measure yields a basis-independent scalar
\begin{equation}
F_2(t)=\frac{1}{L}\sum_{i,j} C^{(\beta)}_{ij;ji}(t)=\frac{1}{L}\,\big|\operatorname{Tr}(y^{2}U_t)\big|^{2} ,
\label{eq:F2_def}
\end{equation}
where $\dim\mathcal{H}_A=d_A$, $\dim\mathcal{H}_B=d_B$, and $L=d_A d_B$. Up to an overall normalization, $F_2(t)$ coincides with the second moment of the spectral form factor at effective inverse temperature $\beta/2$ \cite{Das:2025fcd}.
In our RP numerics we further suppress sample-to-sample fluctuations by an ensemble average,
$\overline{F}_2(t,\beta)\equiv \mathbb E_{H\sim \mathrm{RP}(\gamma)}\,F_2(t,\beta)$.

Motivated by the observation that near $t=0$ the spectrum is real and nonnegative while at later times the Hermitian part develops negative spectral weight, we define the \emph{kernel negativity} as follows. Let
$H_{AB}(t)\equiv (T_{AB}(t)+T^\dagger_{AB}(t))/2$
be the Hermitian part of the spacetime kernel, with eigenvalues $\{\mu_\alpha(t)\}$. We define kernel negativity,  $\mathcal{N}_H(t)$ as
\begin{equation}
\mathcal{N}_H(t)\equiv \sum_{\mu_\alpha(t)<0} |\mu_\alpha(t)|.
\label{eq:neg_def}
\end{equation}
Operationally, $\mathcal N_H(t)$ equals the maximal negative value of $\mathrm{Tr}[H_{AB}(t)M]$ over all POVM effects $0\le M\le \mathbb{I}$, so it quantifies the largest witnessable negative quasiprobability by a single measurement effect.

Equivalently, writing the decomposition $H_{AB}(t)=H_+(t)-H_-(t)$ with $H_\pm(t)\ge 0$ supported on the positive and negative eigenspaces, one has $\mathcal{N}_H(t)=\mathrm{Tr}\,H_-(t)$ and also the trace-norm identity
\begin{equation}
\mathcal{N}_H(t)=\frac{\|H_{AB}(t)\|_1-\mathrm{Tr}\,H_{AB}(t)}{2}
=\frac{\|H_{AB}(t)\|_1-1}{2},
\label{eq:neg_trace_norm_form}
\end{equation}
where $\|X\|_1\equiv \mathrm{Tr}\sqrt{X^\dagger X}$ denotes the trace norm and we used $\mathrm{Tr}\,H_{AB}(t)=\mathrm{Tr}\,T_{AB}(t)=1$.
 The kernel negativity equals the minimal trace-norm change needed to render $H_{AB}(t)$ positive semidefinite, i.e. its trace-norm distance to the positive semidefinite set. Operationally, this is the smallest amount of negative quasiprobability encoded in $H_{AB}(t)$, in the sense that it quantifies how much negative spectral weight must be removed to obtain a positive semidefinite (PSD) density operator.

Finally, we define the time-averaged kernel negativity over a window $\mathcal{W}$ for a fixed RP parameter $\gamma$,
\begin{equation}
\overline{\mathcal{N}}_H(\gamma)\equiv \frac{1}{|\mathcal{W}|}\sum_{t\in\mathcal{W}} \mathcal{N}_H\!\left(T_{AB}(t;\gamma)\right),
\label{eq:neg_time_avg}
\end{equation}
which we use as a phase-sensitive dynamical diagnostic alongside $Z_2(t)$, $\|T_{AB}(t)\|_2^2$, and $\mathcal{I}_2(t)$.

\paragraph*{Spacetime density kernel and fractality---} 
Figure~\ref{fig:1} summarizes the response of the measures that we just defined to the Rosenzweig--Porter control parameter $\gamma$. We scan $\gamma\in[0.1,3]$ in steps of $0.1$, color-coded by a rainbow map, with small $\gamma$ at the warm end and large $\gamma$ at the cool end, and monitor three complementary diagnostics: the second Tsallis entropy, the Frobenius  norm of $T_{AB}(t)$, and the $2$-imagitivity, which controls causal influence via commutators.

Here, we present results for $M=6$ qubits with a symmetric bipartition, where subsystem A and B consist of the first and the last three qubits, respectively. We have checked that other bipartitions exhibit qualitatively similar dynamical behavior; these results are summarized in Supplemental Material Sec.~\ref{sm2} and Fig.~\ref{fig:3}. Panel~\ref{fig:ttbar} shows the Frobenius norm of the spacetime kernel. $\|T_{AB}(t)\|_2^2$ is the squared norm of the singular-value spectrum, it acts as a ``temporal purity'' of the spacetime kernel. Numerically we observe a rapid approach to a $\gamma$-dependent late-time plateau shown in Fig.~\ref{fig:ttbar}, with the ergodic regime exhibiting the highest plateau and the localized regime the lowest. Interpreted in the RP language, this is consistent with the idea that stronger eigenvector ergodicity produces a more effective randomization of amplitudes across the $A|B$ partitioning of basis indices, which in turn yields a larger steady-state kernel weight measured in this Schatten-2 metric. The fractal phase falls in between, indicating intermediate speed of exploration and saturation value in the Hilbert space. 

 To discuss the SFF-like behavior in panel~\ref{fig:tsq}, we benchmark the real part of the second Tsallis entropy, against the coarse-grained two-point function $\overline{F}_2(t,\beta)$ defined in \eqref{eq:F2_def} and shown in~\ref{fig:f2bar}, evaluated here at infinite temperature ($\beta=0$). At $\beta=0$, $\overline{F}_2(t,0)$ is an ensemble-averaged version of the spectral form factor and is therefore directly sensitive to long-range spectral correlations (level repulsion and rigidity). Fig.~\ref{fig:tsq} and \ref{fig:f2bar} compare $\mathrm{Re}\,Z_2(t)$ (top row) to $\overline{F}_2(t,0)$ (bottom row) across the RP phase diagram. In the ergodic window ($0<\gamma\lesssim 1$), both quantities exhibit the canonical dip--ramp--plateau profile: the early-time dip reflects level repulsion, the ramp encodes the build-up of spectral rigidity, and the plateau sets by the finite Hilbert-space dimension. In the fractal regime ($1\lesssim\gamma\lesssim 2$), the same structure persists but is systematically weakened: the ramp is reduced and stretched over time, consistent with an intermediate degree of spectral rigidity and a reduced effective number of interfering channels set by the non-ergodic support of eigenvectors. Deep in the localized regime ($\gamma\gtrsim 2$), the ramp is strongly suppressed and the curves approach a plateau more directly, as expected when level correlations become Poisson-like and long-range rigidity is lost. Together, Figs.~\ref{fig:tsq} and \ref{fig:f2bar} show that timelike-kernel observables simultaneously diagnose (i) dynamical non-Hermiticity and (ii) spectral rigidity, and that both collapse in a controlled manner across the ergodic to fractal to localized crossover. A phase-resolved presentation of this comparison on a logarithmic time axis is given in Supplemental Material, Sec.~\ref{sm1} and Fig.~\ref{fig:2}.

 Panel~\ref{fig:schntab} plots $\mathcal{I}_2(t)$ defined in \eqref{eq:imag_def}. By construction, $\mathcal{I}_2(t)$ vanishes for Hermitian kernels and grows when time ordering becomes operationally distinguishable. In the construction of \cite{Milekhin:2025ycm}, nonzero $\mathcal{I}_2$ is a robust indicator of causal influence as it upper-bounds commutators and vanishes if and only if signaling is impossible. In our RP numerics, $\mathcal{I}_2(t)$ grows more rapidly and saturates to a higher plateau for small $\gamma$, while it is strongly suppressed for large $\gamma$, with the fractal regime again interpolating. This is a direct dynamical signature of eigenvector ergodicity in the computational basis: when RP eigenvectors are delocalized, the unitary $U(t)$ has many comparable matrix elements in that basis, producing strong $A\rightarrow B$ operator mixing and hence a large separation between $T_{AB}(t)$ and $T_{AB}^\dagger(t)$; when eigenvectors are localized, $U(t)$ is approximately sparse/diagonal, and $T_{AB}(t)$ remains closer to an effectively classical PSD-like object.

\paragraph{Kernel negativity and quasiprobability content---} Panels~\ref{fig:negtab} and ~\ref{fig:D2_ttbar_ntavg} introduce and summarize our new diagnostic: the \emph{kernel negativity} $\mathcal{N}_H(t)$, defined in \eqref{eq:neg_def}. Numerically, panel~\ref{fig:negtab} shows that $\mathcal{N}_H(t)$ is near zero at very early times when the kernel behaves most like an ordinary reduced density matrix with a real nonnegative spectrum, and then becomes finite as time evolution generates noncommuting, interference-dominated temporal correlations; crucially, the ergodic regime develops the largest and fastest-growing negativity, while the localized regime remains close to zero over the same window. We find that the approach to the late-time plateau for both $\|T_{AB}(t)\|_2^2$ and $\mathcal N_H(t)$ is well fit by a Kohlrausch--Williams--Watts (stretched/compressed exponential) function \cite{Kohlrausch1854, Williams1970}; fit details and extracted parameters are reported in the Supplemental Material Sec.~\ref{sm2}. The existence of the Kohlrausch--Williams--Watts type function is deeply connected to a sub-diffusive physics that is known to exist within the multifractal phase of the RP-model, see {\it e.g.}~\cite{Khaymovich_2021}. Usually, this sub-diffusive dynamics is observed generically within the return amplitude. Our results, however,  further indicate a wider ubiquity of the exponential responses in this class of models, albeit compressed rather than stretched.

Panel~\ref{fig:D2_ttbar_ntavg} summarizes the phase dependence of two time-averaged kernel observables, the negativity $\overline{\mathcal N}_H(\gamma)$ and the Frobenius norm $\overline{\|T_{AB}\|_2^2}(\gamma)$ averaged over the window $W=[0,50]$, together with the fractal dimension $D_2(\gamma)$ extracted independently from inverse-participation-ratio scaling. To facilitate a direct comparison, the time-averaged kernel negativity and Frobenius norm are rescaled according to
\begin{equation}
\widetilde{X}(\gamma)=\frac{X(\gamma)-X(\gamma,t{=}0)}{\max_{\gamma}[X(\gamma)-X(\gamma,t{=}0)]},
\label{eq:rescale}
\end{equation}
where $X\in\{\overline{\mathcal N}_H,\overline{\|T_{AB}\|_2^2}\}$. This normalization maps both quantities to the interval $[0,1]$, removes nonuniversal offsets, and highlights their relative variation across the phase diagram. Both rescaled kernel observables decrease monotonically as $\gamma$ is tuned from the ergodic to the localized regime, with clear changes in slope near the conventional RP boundaries at $\gamma\simeq 1$ and $\gamma\simeq 2$. Strikingly, their trends closely track the known behavior of the fractal dimension $D_2(\gamma)$, which takes the values $D_2\simeq 1$ in the ergodic phase, $0<D_2<1$ in the fractal phase, and $D_2\simeq 0$ deep in the localized regime, as established in the RP literature. While $D_2$ is a static property of eigenstates, the quantities $\overline{\mathcal N}_H$ and $\overline{\|T_{AB}\|_2^2}$ are dynamical and extracted from real-time evolution.

This empirical alignment supports a unified physical picture: both the time-averaged kernel negativity and the Frobenius norm quantify the extent to which the dynamics can generate interference among a large number of basis states under the chosen bipartition. In the RP ensemble, the number of effective interference channels is controlled by the eigenvector support size, which is precisely what the fractal dimension $D_2$ measures. From this perspective, $\overline{\mathcal N}_H$ and $\overline{\|T_{AB}\|_2^2}$ act as basis-sensitive, operationally meaningful dynamical witnesses of non-ergodic extended structure, complementing imagitivity and providing a time-domain counterpart to the fractal phase.

\paragraph*{Discussion and outlook---} In this Letter we used the spacetime density kernel framework to extract simultaneous dynamical signatures of (i) eigenvector ergodicity/fractality and (ii) spectral rigidity in the Rosenzweig--Porter ensemble, using distinct qubit bipartitions. Several directions follow naturally. The operational meaning of kernel negativity suggests experimental and algorithmic routes. Precise measurement protocols for the observables studied here are an important direction for future work. To that end, \cite{Milekhin:2025ycm} already provides us with concrete measurement protocols that can further be generalized to measure, {\it e.g.}~kernel negativity, among the other observables, which we have explored here.  The timelike entanglement is not expected to be monotone under arbitrary LOCC in general, but it would be valuable to identify approximate, constrained, or operational settings in which kernel-based quantities admit monotonicity or resource-theoretic interpretations.

Motivated by the ubiquity of the RP model in capturing a smooth transition between an integrable and a chaotic dynamical system, through a non-ergodic multi-fractal phase, it is natural to export these diagnostics beyond RP. Especially to the random-matrix universality class, to many-body Hamiltonians with mobility edges or many-body localization proximity, Floquet systems, and to holographic/field-theoretic settings where timelike entanglement was originally motivated to test whether kernel negativity and imagitivity provide a broadly applicable, resource-like characterization of dynamical chaos and non-ergodic extended phases. For example, it is likely that the kernel is sensitive to the Bohigas--Giannoni--Schmit conjecture in the context of the RMT-universality.

Within continuum quantum field theory, and especially in holographic settings, it will be interesting to understand what dynamical information is encoded in the spacetime kernel. The emergence of space-time is connected to quantum entanglement \cite{VanRaamsdonk:2009ar, VanRaamsdonk:2010pw}, and, in particular, the emergence of time is associated with a notion of the timelike entanglement~\cite{Doi_2023}. Considering a black hole in a holographic space-time, this is directly connected to the physics inside the event horizon, a region that is classically inaccessible to the asymptotic observer. Our results are particularly encouraging from a broad and qualitative perspective: the presence of a classical event horizon is deeply connected with the presence of a highly chaotic dynamics, both semi-classically \cite{Maldacena_2016} as well as quantum gravitationally \cite{Cotler_2017}. The spacetime kernel encodes dynamical information across a broad range of time scales and, as we show here, resolves chaos-to-integrable crossovers. This motivates the question of whether analogous kernel diagnostics could provide an effective, coarse-grained probe of horizon formation and interior physics in tractable models. At the very least, it is immediately expected to detect the existence of Hawking-Page like phase transitions, through the implicit relation between the kernel and operator complexity along the lines of \cite{Kundu:2023hbk}. We hope to address some of these aspects in future.

\begin{acknowledgments}
\vspace{2em}
\noindent The authors would like to thank Taishi Kawamoto, Alexey Milekhin, Krishanu Roychowdhury, Ashish Shukla and Tadashi Takayanagi for useful discussions and comments. A.K. acknowledges the support of the Humboldt Research Fellowship for Experienced Researchers by the Alexander von Humboldt
Foundation and for the hospitality of Theoretical Physics III, Department of Physics and Astronomy, Julius-Maximilians-Universit\"{a}t W\"{u}rzburg and the support from the ICTP through the Associates Programme (2024-2030) during the course of this work. R.N.D.'s work leading to this publication was supported by the PRIME programme of the German Academic Exchange Service (DAAD) with funds from the German Federal Ministry of Research, Technology and Space (BMFTR).  R.N.D. is also supported by Germany's Excellence Strategy through the W\"urzburg‐Dresden Cluster of Excellence on Complexity, Topology and Dynamics in Quantum Matter ‐ ctd.qmat (EXC 2147, project‐id 390858490).
\end{acknowledgments}

\newpage
~
\newpage
\onecolumngrid
\renewcommand{\theequation}{A.\arabic{equation}}
\setcounter{equation}{0}

\section*{Supplemental Material}
\section{Ensemble-averaged two-point function and the second Tsallis entropy}
\label{sm1}
\begin{figure*}[hbtp]
    \centering
    
     \begin{subfigure}[b]{\textwidth}
         \centering
         \includegraphics[width=\textwidth]{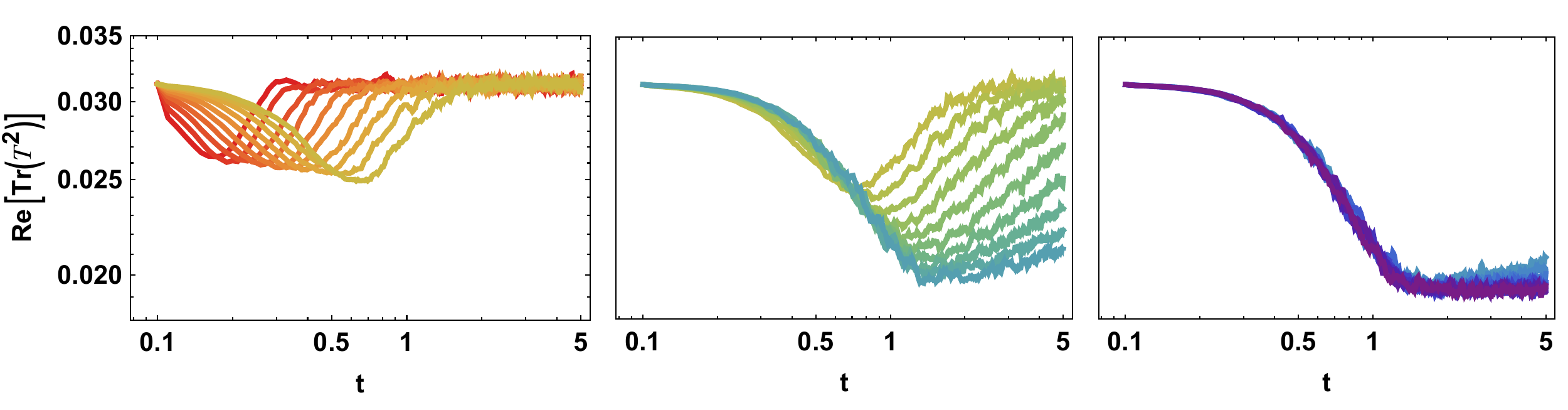}
         \caption{}
         \label{fig:trtsqres}
     \end{subfigure}
     \\
     \begin{subfigure}[b]{\textwidth}
     \centering
         \includegraphics[width=\textwidth]{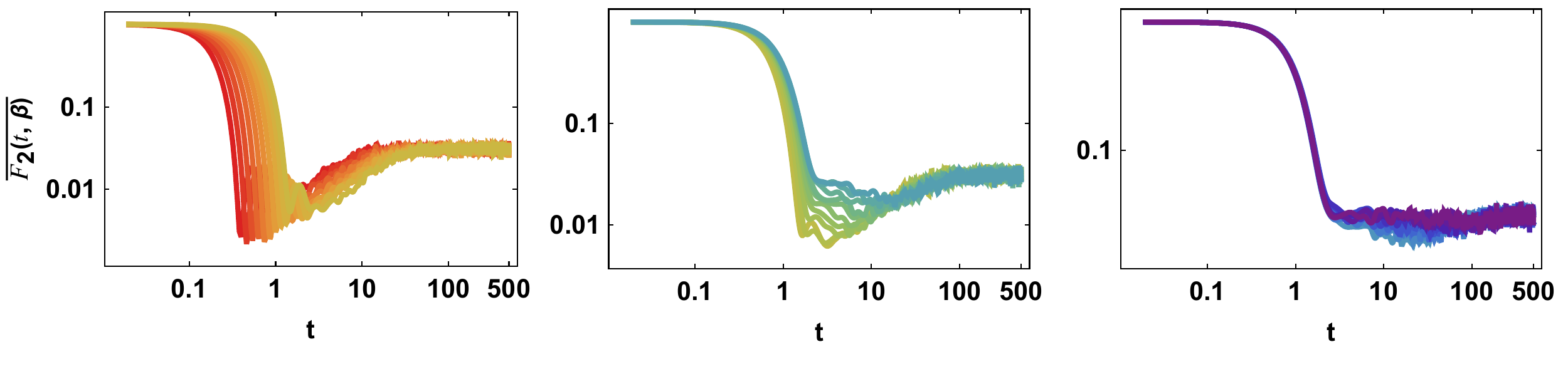}
         \caption{}
         \label{fig:f2barres}
     \end{subfigure}
     
 \caption{Top row: second timelike Tsallis entropy, $\mathrm{Re}\,\mathrm{Tr}[T_{AB}(t)^2]$.
Bottom row: averaged two-point correlator $\overline{F}_2(t,\beta)$ versus $t$.
Each curve corresponds to a RP parameter $\gamma$ (step $0.1$), with the same rainbow color scheme as Fig.~\ref{fig:1} (warm: small $\gamma$, cool: large $\gamma$); the time axis is logarithmic.
Left column ($0.1\le \gamma\le 1$, ergodic): clear dip--ramp--plateau in both quantities.
Middle ($1\le \gamma\le 2$, fractal): ramp weakens and becomes $\gamma$ dependent, with a delayed approach to the plateau.
Right ($2\le \gamma\le 3$, localized): ramp is strongly suppressed, and both curves approach the plateau directly.
Overall, $\mathrm{Re}\,\mathrm{Tr}[T_{AB}(t)^2]$ displays comparable dynamics to SFF by reproducing the dip--ramp--plateau across the RP phase diagram. The colour scheme used here is the same as in Fig.~\ref{fig:1}.}

\label{fig:2}
\end{figure*}

Fig.~\ref{fig:2} benchmarks the real part of the second Tsallis entropy against a standard spectral diagnostic built from an ensemble-averaged two-point function. The bottom row shows the coarse-grained correlator $\overline{F}_2(t,\beta)$ introduced in Ref.~\cite{Das:2025fcd}, evaluated at infinite temperature ($\beta=0$), where it reduces to a smoothed/ensemble-averaged version of the spectral form factor and is therefore directly sensitive to long-range spectral correlations (spectral rigidity). The top row shows the corresponding kernel quantity $\mathrm{Re}\,\mathrm{Tr}[T_{AB}(t)^2]$, computed from the two-leg spacetime density kernel $T_{AB}(t)$ defined by the trace-pairing relation in \eqref{eq:tab_basis_recon} of the main text and using the same time window and sampling as in Fig.~\ref{fig:1}. The three columns separate the RP control parameter into the conventional phase intervals: $0<\gamma\le 1$ (ergodic), $1\le\gamma\le 2$ (fractal/non-ergodic extended), and $2\le\gamma\le 3$ (localized).

The key observation is that $\mathrm{Re}\,\mathrm{Tr}[T_{AB}(t)^2]$ reproduces, within the same dynamical framework that underlies our entanglement and non-Hermiticity diagnostics, the characteristic dip--ramp--plateau phenomenology of $\overline{F}_2(t,0)$. In the ergodic regime (left column), both quantities exhibit a pronounced early-time dip (the ``correlation hole'') followed by a clear ramp and a late-time plateau, consistent with Wigner--Dyson level repulsion and robust spectral rigidity. In the fractal regime (middle column), the same structure persists but is weakened and stretched: the ramp becomes less sharp and more $\gamma$ dependent, indicating intermediate spectral rigidity in the non-ergodic extended phase. In the localized regime (right column), the ramp is strongly suppressed and the curves approach the plateau more directly, consistent with the loss of long-range correlations and the emergence of Poisson-like spectral statistics.

Physically, the agreement between the two rows can be understood as follows. The averaged two-point function $\overline{F}_2(t,0)$ isolates spectral correlations of the RP Hamiltonian itself, while $\mathrm{Re}\,\mathrm{Tr}[T_{AB}(t)^2]$ packages interference between the forward and backward operator orderings encoded in the spacetime kernel. In the chaotic window ($\gamma\lesssim 2$), eigenvector ergodicity generates many comparable-magnitude contributions with effectively random phases, producing the cancellations responsible for the dip and the gradual buildup of correlations responsible for the ramp. When $\gamma>2$, eigenvectors become localized in the computational basis, the available interference channels collapse, and both the spectral rigidity and the kernel-induced ramp disappear. Thus Fig.~\ref{fig:2} shows that the real part of the second Tsallis entropy $\mathrm{Re}\,\mathrm{Tr}[T_{AB}(t)^2]$ provides a kernel-native, timelike analogue of a spectral form factor diagnostic, complementing Fig.~\ref{fig:1} by demonstrating that the same framework captures both eigenvector (non-)ergodicity and spectral chaos.

 \section{Dependence of spacetime-kernel observables on the choice of bipartition}

\label{sm2}

\begin{figure*}[hbtp]
    \centering
     \begin{subfigure}[b]{0.32\textwidth}
     \centering
         \includegraphics[width=\textwidth]{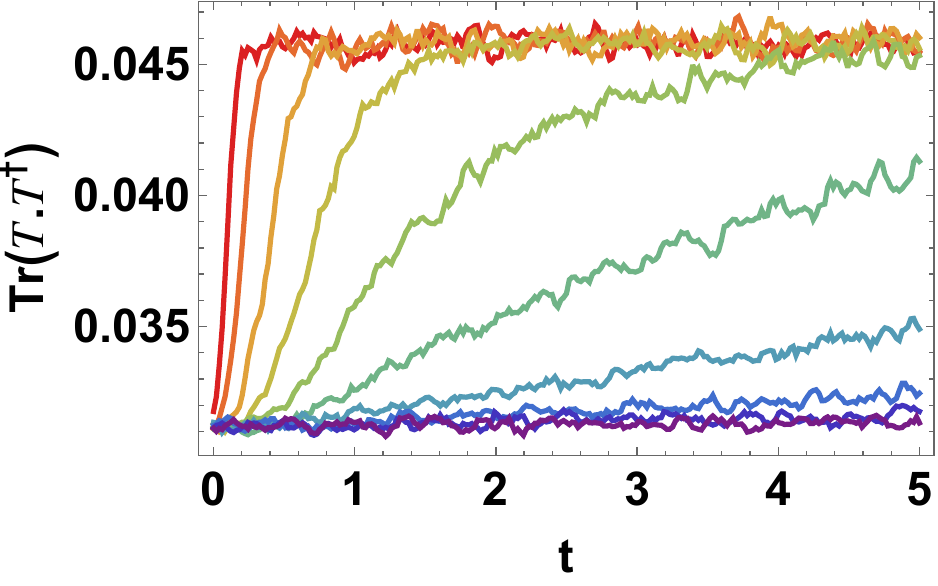}
         \caption{$A=\{1,2\}$, $B=\{3,4,5,6\}$ (disjoint, unequal).}
         \label{fig:12 _ 3456 _ttbar}
     \end{subfigure}
     \hfill
     \begin{subfigure}[b]{0.32\textwidth}
         \centering
         \includegraphics[width=\textwidth]{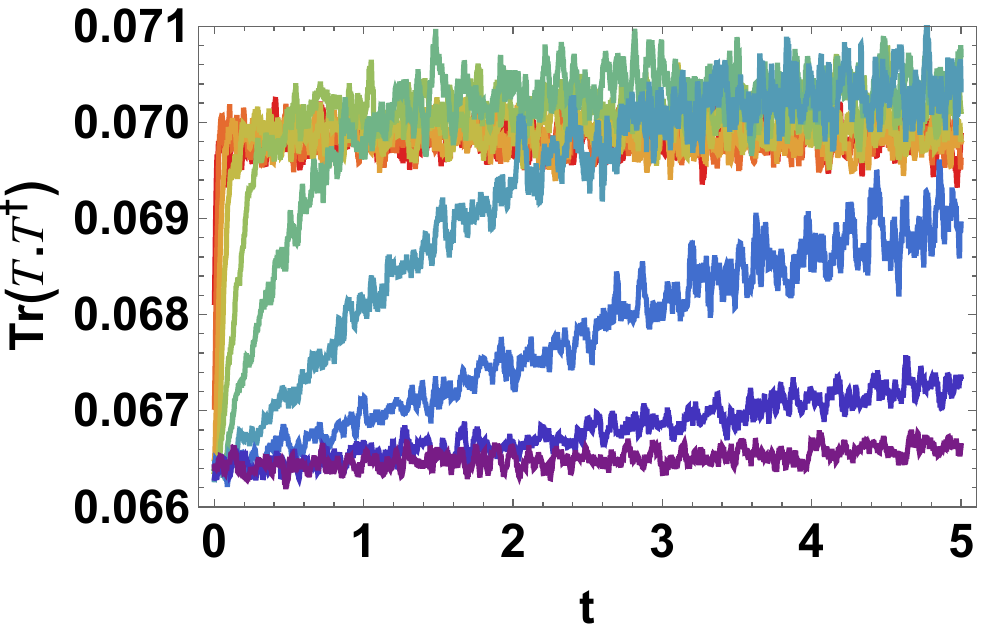}
         \caption{$A=\{1,2\}$, $B=\{5,6\}$ (disjoint, separated).}
         \label{fig:12 _ 56 _ttbar}
     \end{subfigure}
     \hfill
     \begin{subfigure}[b]{0.32\textwidth}
         \centering
         \includegraphics[width=\textwidth]{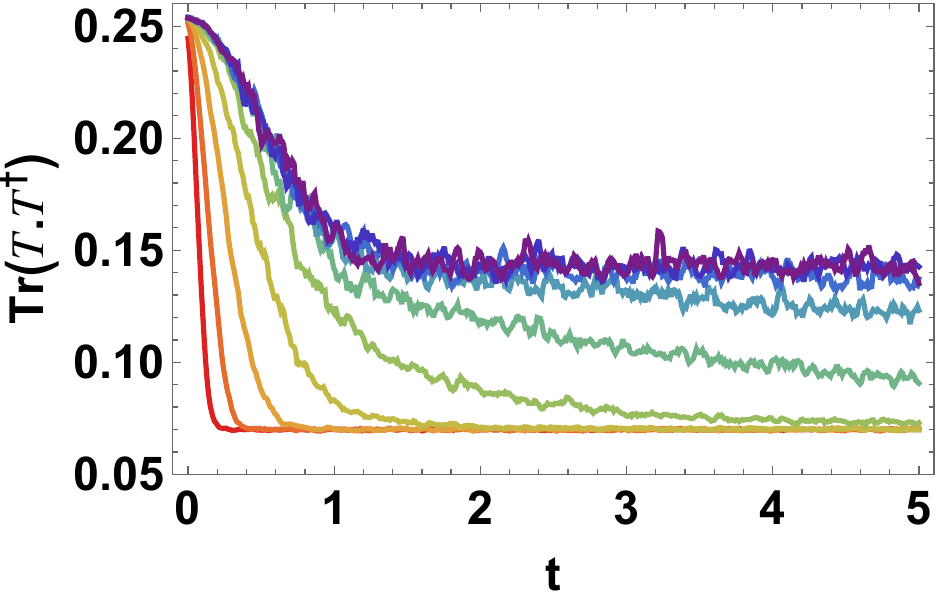}
         \caption{$A=\{1,2\}$, $B=\{2,3\}$ (overlapping).}
         \label{fig:12 _ 23 _ttbar}
     \end{subfigure}
     \\
     \begin{subfigure}[b]{0.32\textwidth}
     \centering
         \includegraphics[width=\textwidth]{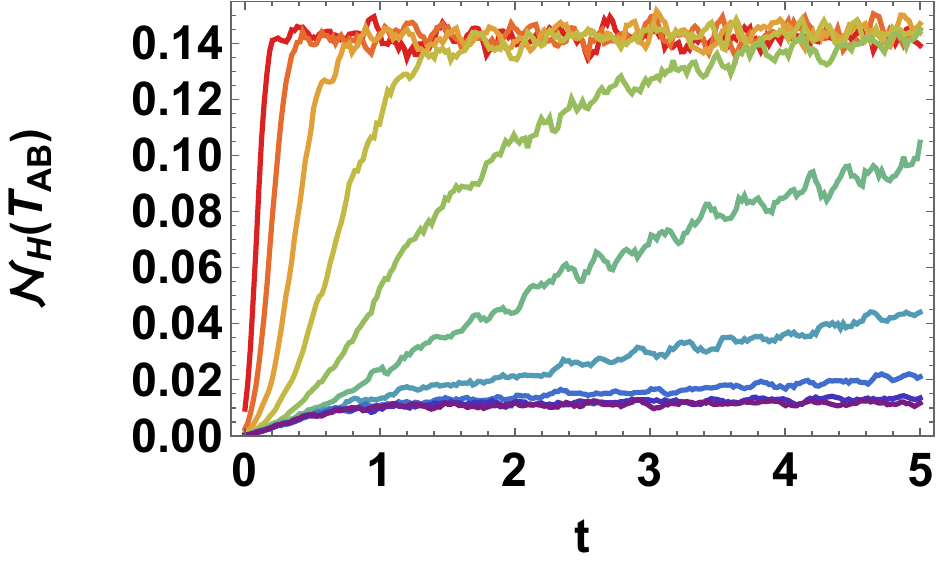}
         \caption{$A=\{1,2\}$, $B=\{3,4,5,6\}$.}
         \label{fig:12 _ 3456 _nh}
     \end{subfigure}
     \hfill
     \begin{subfigure}[b]{0.32\textwidth}
     \centering
         \includegraphics[width=\textwidth]{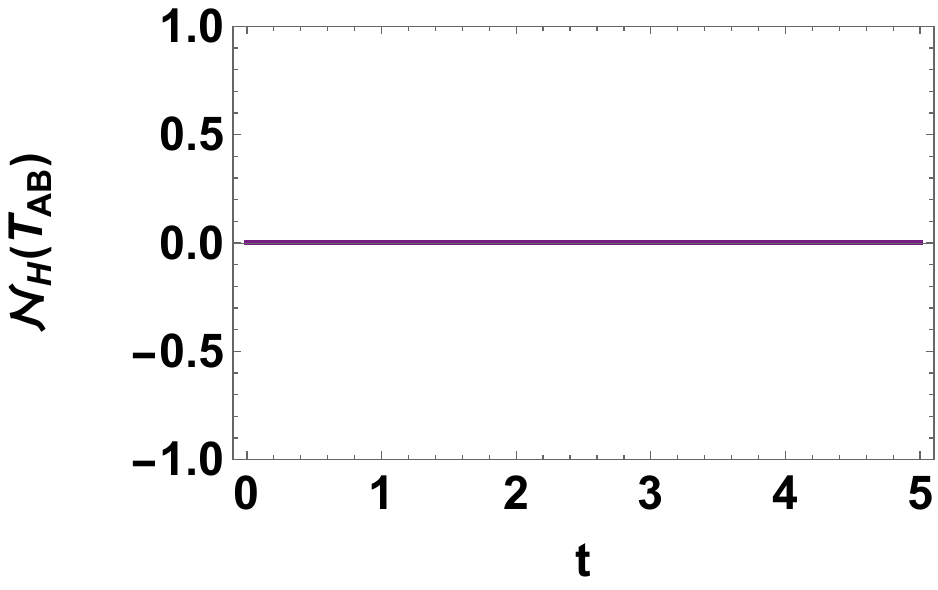}
         \caption{$A=\{1,2\}$, $B=\{5,6\}$.}
         \label{fig:12 _ 56_nh}
     \end{subfigure}
     \hfill
     \begin{subfigure}[b]{0.32\textwidth}
         \centering
         \includegraphics[width=\textwidth]{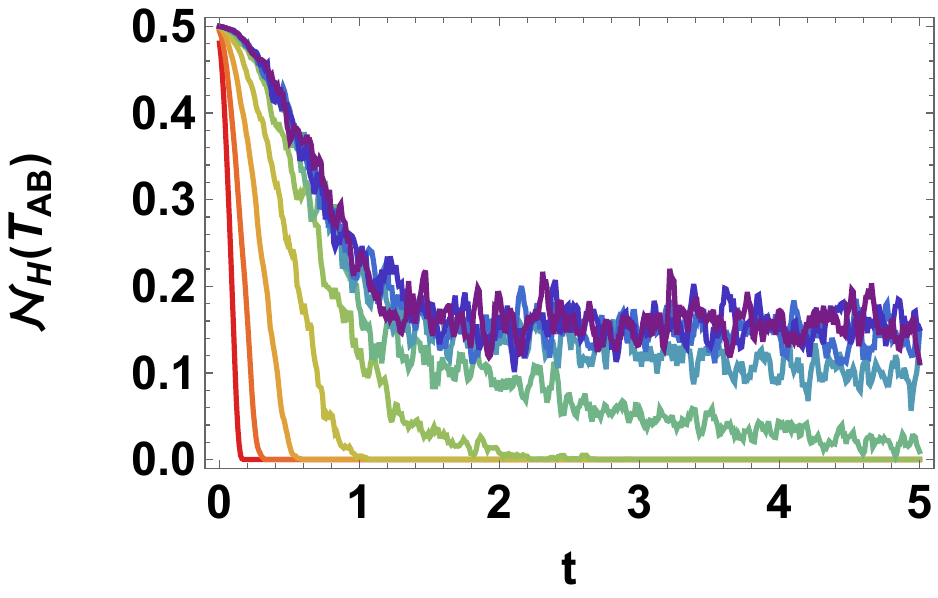}
         \caption{$A=\{1,2\}$, $B=\{2,3\}$.}
         \label{fig:12 _ 23_nh}
     \end{subfigure}
     
    \caption{
    Top row (a--c): Frobenius norm $\|T_{AB}(t)\|_2^2$ for three bipartitions of a six-qubit RP system. In all cases, $\|T_{AB}(t)\|_2^2$ grows more rapidly and saturates to a higher plateau for small $\gamma$ (ergodic), while both the growth and plateau are suppressed at large $\gamma$ (localized), with the fractal regime interpolating between these limits.
    Bottom row (d--f): kernel negativity $\mathcal N_H(t)$ for the same bipartitions. For the disjoint partitions (d), negativity is generated dynamically and saturates to a $\gamma$-dependent value that decreases with increasing $\gamma$; the separated disjoint partition (e) exhibits the no growth over the time window. For the overlapping partition (f), $\mathcal N_H(t)$ is nonzero already at $t=0$ and relaxes to a late-time value that again decreases with increasing $\gamma$. Overall, the phase-sensitive trends of both $\|T_{AB}(t)\|_2^2$ and $\mathcal N_H(t)$ are robust, while their dynamical behavior and absolute magnitude depend on whether the subsystems overlap. The colour scheme used here is the same as in Fig.~\ref{fig:1}. }
    \label{fig:3}
\end{figure*}

The spacetime density kernel $T_{AB}(t)$ is defined once a choice of subsystems $A$ and $B$ is specified. While the main text focuses on an equal, disjoint bipartition of the six-qubit Hilbert space, it is important to verify that the phase-sensitive signatures we report are not an artifact of this particular cut. Fig.~\ref{fig:3} compares alternatives and illustrates which features are robust and which depend on specific details of the chosen supports.

We consider three bipartitions with subsystem $A$ fixed to the first two qubits, $A=\{1,2\}$, and varying $B$:
(i) a complementary, disjoint cut $B=\{3,4,5,6\}$,
(ii) a disjoint but non-complementary choice $B=\{5,6\}$, leaving a spectator set $C=\{3,4\}$ that is traced over in the correlator defining $T_{AB}(t)$,
and (iii) an overlapping choice $B=\{2,3\}$ that shares a qubit with $A$.
In all panels, curves are color-coded by the RP parameter $\gamma$ (warm colors: smaller $\gamma$; cool colors: larger $\gamma$), with the same convention as in Fig.~\ref{fig:1}.

Panels~(\ref{fig:12 _ 3456 _ttbar}--\ref{fig:12 _ 23 _ttbar}) show the Frobenius norm $\|T_{AB}(t)\|_2^2=\mathrm{Tr}[T_{AB}(t)T_{AB}^\dagger(t)]$ for the three bipartitions. The qualitative phase dependence is stable across all choices: small $\gamma$ (ergodic) exhibits faster early-time growth and a higher late-time plateau, large $\gamma$ (localized) is strongly suppressed in both growth and saturation, and the intermediate regime interpolates smoothly.
This robustness is expected because $\|T_{AB}\|_2^2$ is primarily controlled by how effectively the unitary dynamics mixes operator weight between the chosen $A$- and $B$-subsystems, which in the RP ensemble is governed by eigenvector support in the computational basis rather than by any underlying spatial geometry.

At the same time, the absolute scale and fine structure of $\|T_{AB}\|_2^2$ do depend on the cut. Changing the sizes of $A$ and $B$ changes the dimension of the kernel’s operator space and hence the typical baseline and plateau values. In particular, the complementary cut in \ref{fig:12 _ 3456 _ttbar} samples correlations between a small $A$ and a large $B$, while \ref{fig:12 _ 56 _ttbar} probes two small, disjoint subsystems embedded in a larger environment; these choices differ quantitatively but preserve the same ordering of ergodic versus non-ergodic behavior.

Panels~(\ref{fig:12 _ 23 _ttbar}--\ref{fig:12 _ 23_nh}) show the kernel negativity $\mathcal N_H(t)$, defined from the Hermitian part
$H_{AB}(t)\equiv [T_{AB}(t)+T_{AB}^\dagger(t)]/2$
as the total negative spectral weight of $H_{AB}(t)$.
This observable is more sensitive to the detailed subsystem choice because it detects genuinely quasi-probabilistic structure: negative eigenvalues of $H_{AB}(t)$ require sufficiently strong destructive interference among the operator channels contributing to the two-time correlator.

For the disjoint complementary partition $A=\{1,2\}$, $B=\{3,4,5,6\}$ (panel d), $\mathcal N_H(t)$ is initially small and is generated dynamically, approaching a $\gamma$-dependent saturation value. This is the cleanest setting conceptually: at $t=0$ the kernel reduces to a reduced state in the $A\otimes B$ description, while at $t>0$ the dynamics produces noncommuting temporal correlations that can drive $H_{AB}(t)$ away from positivity.

For the disjoint but non-complementary choice $A=\{1,2\}$, $B=\{5,6\}$ (panel e), $\mathcal N_H(t)$ exhibits essentially no growth over the same time window. A natural interpretation is that tracing over the spectator qubits $C=\{3,4\}$ introduces additional coarse graining that suppresses the negative spectral weight of the Hermitian part, even though $\|T_{AB}(t)\|_2^2$ still displays clear phase-dependent growth and saturation. In other words, this cut retains sensitivity to eigenvector ergodicity through norm growth, but it can wash out the quasiprobability signature captured by $\mathcal N_H(t)$.

The overlapping choice $A=\{1,2\}$ and $B=\{2,3\}$ (panel \ref{fig:12 _ 23_nh}) behaves qualitatively differently: $\mathcal N_H(t)$ is already nonzero at $t=0$ and relaxes toward a late-time value that again decreases with increasing $\gamma$. The origin is structural. When $A$ and $B$ share degrees of freedom, the bilinear form defining $T_{AB}(t)$ couples probes that act on the same qubit(s), so even at $t=0$ the kernel need not resemble a reduced density operator on a tensor product $A\otimes B$; consequently, positivity of the Hermitian part is not guaranteed at the initial time. Time evolution then redistributes this initial non-positivity and drives the kernel toward a $\gamma$-dependent steady regime. We also verified the same qualitative ordering for $M=4,5,7,8$ with random pure state.

Taken together, Fig.~\ref{fig:3} supports two practical conclusions. First, the phase-sensitive trends of Schatten-type norms are robust across reasonable bipartitions, reinforcing their role as general dynamical probes of the RP ergodic--fractal--localized structure. Second, kernel negativity is an especially sharp diagnostic when the subsystems are disjoint, whereas overlapping or disjoint and separated subsystem choices can modify the early-time baseline and substantially reduce the dynamical negativity signal. These observations motivate our main-text focus on disjoint, balanced bipartitions, for which the $t=0$ kernel is closest in spirit to an ordinary reduced state and the subsequent emergence of non-positivity can be unambiguously attributed to real-time dynamics.

Across system sizes and bipartitions, both the kernel negativity $\mathcal{N}_H(t)$ and the spacetime-kernel purity
$\mathrm{Tr}\!\left[T_{AB}(t)T_{AB}^\dagger(t)\right]$
are well captured by a Kohlrausch--Williams--Watts (KWW) function,
\begin{equation}
\label{eq:KohlrauschFit}
Q(t)\;\approx\;Q_\infty+\bigl(Q_0-Q_\infty\bigr)\exp\!\Big[-(t/\tau)^{\beta}\Big],
\qquad Q\in\Big\{\mathcal{N}_H,\;\mathrm{Tr}[T_{AB}T_{AB}^\dagger]\Big\}.
\end{equation}
Here $Q_0\equiv Q(t=0^+)$ and $Q_\infty$ denote the early-time baseline and late-time plateau, $\tau$ sets the
characteristic timescale, and $\beta$ controls the stretching/compression. In all fits we find $\beta>1$,
corresponding to a compressed exponential. In our case, $\tau$ is $\mathcal{O}(1)$ and increases with $\gamma$.

For an increasing observable it is convenient to rewrite \eqref{eq:KohlrauschFit} as
$f(t)=A\!\left(1-e^{-(t/\tau)^{\beta}}\right)$ where $A\equiv Q_\infty-Q_0$ is the total rise from the baseline to the plateau.

At early times,
\begin{equation}
\label{eq:EarlyTimeExpansion}
f(t)=A\Big[(t/\tau)^{\beta}-\tfrac12(t/\tau)^{2\beta}+\cdots\Big]
\;\approx\;A\,(t/\tau)^{\beta},
\end{equation}
so the initial growth is polynomial of degree $\beta$.
At late times the deviation from the plateau obeys
\begin{equation}
\label{eq:LateTimeTail}
A-f(t)=A\,e^{-(t/\tau)^{\beta}},
\end{equation}
i.e.\ the approach to the asymptote is exponential in the stretched variable $(t/\tau)^{\beta}$ and therefore faster than a simple exponential in $t$ when $\beta>1$.
Equation~\eqref{eq:KohlrauschFit} therefore describes both decays and growth, depending on the sign of $(Q_0-Q_\infty)$.

\bibliography{ref}
\end{document}